\documentclass[12pt]{article}
\usepackage{graphicx}
\usepackage{epsfig}
\usepackage{cite}
\usepackage{amsmath,amssymb}
\begin{document}

\hspace*{\fill}{\small {published in: Mod. Phys. Lett. A 29 (2014) 1450194}}
\vspace{0.5cm}

\begin{center}
{\bfseries {\large THE QCD ANALYSIS OF THE COMBINED SET FOR THE
$F_3$ STRUCTURE FUNCTION DATA BASED ON THE ANALYTIC APPROACH}}
\vskip 5mm {A.~V. Sidorov$^{\dag}$ and O.~P.
Solovtsova$^{\dag,\ddag}$} \vskip 5mm
{\small {\it $^\dag$ Joint Institute for Nuclear Research, 141980 Dubna, Russia}} \\
{\small {\it $^\ddag$ Gomel State Technical University, 246746
Gomel, Belarus}}
\end{center}

\begin{abstract}
We apply analytic perturbation theory to the QCD analysis of the
$xF_3(x,Q^2)$ structure function considering a combined set of deep
inelastic scattering data presented by several collaborations, and
extract values of the scale parameter $\Lambda_{QCD}$, the
parameters of the form of the $xF_3$ structure function, and the
$x$-shape of the higher twist contribution. We study the difference
between the results obtained within the standard perturbative and
analytic approaches in comparison with the experimental errors and
state that the greatest difference occurs for large $x$. \\



\end{abstract}

\section{Introduction}
High-precision measurements of the deep inelastic scattering data
open a possibility for testing new theoretical ideas in a wide
kinematic range of $x$ and $Q^2$. In the present paper, we report
the results of the application of the analytic approach in QCD
\cite{apt96-7-1,apt96-7-2}, called the analytic perturbation theory
(APT), to the QCD analysis of the experimental data on the $xF_3$
structure function. In contrast to the perturbative running coupling
which is spoiled by unphysical singularities in the infrared region
at a scale $Q \sim \Lambda_{QCD}$, in the framework of the APT, the
analytic coupling has no unphysical singularities. This fact
stimulated applications of the APT and its modifications supporting
correct analytic properties for various physical processes (see
Refs.~\cite{ShSol06,Baldicchi:2007ic,Bakulev:2008td,Cvetic:2009kw,Cvetic:2011ym,Cvetic:2012pw,Khandramai:2011zd,Ayala:2012yw,Allendes:2014fua}),
especially after the generalization of the APT method to the
noninteger powers of the running coupling \cite{BMS05,BMS06}.

This paper is a continuation of our previous analysis \cite{SS-F3-1}
of the $xF_3$ structure function data of the CCFR collaboration
\cite{CCFR97}. In the present analysis, we include in the
consideration a set of various experimental $xF_3$ data: the CDHS
\cite{CDHS}, SCAT \cite{SCAT}, BEBC-GARGAMELLE
\cite{BEBCGARGAMELLE}, BEBC-WA59 \cite{BEBC-WA59}, NuTeV
\cite{NuTeV} and CHORUS \cite{CHORUS}. Note that similar
perturbative analysis was performed earlier in
Ref.~\cite{Sidorov-96}, but later the new data of the NuTeV and
CHORUS collaborations were published, and we include these data in
the present analysis. In accordance with the result of
Ref.~\cite{CDHS} concerning  the disagreement of CDHS data with
perturbative QCD at small $x$, a cut $x\geq0.35$ was applied to
their data. The total number of experimental points is 289 and
considerably exceeds the number of points considered before in
Ref.~\cite{SS-F3-1}. It allows us to determine parameters of the
form of the $xF_3$ structure function more precisely.
The kinematic region of a combined set of data is $0.015<x<0.8$ and
$ 0.5~{\mbox{\rm GeV}}^2 <Q^2< 196$ GeV$^2$.
Thus, in addition to the previous analysis in Ref.~\cite{SS-F3-1}
the kinematic region $ 0.5~{\mbox{\rm GeV}}^2 <Q^2< 1.3$ GeV$^2$ is
now included. This region provides a possibility for more precise
determination of the parameter $\Lambda_{QCD}$ and the higher twist
(HT) contribution. In order to achieve a more direct comparison of
the perturbation theory (PT) and APT results for fit of a combine
set of data, we do not introduce the normalization factors for the
data of an individual collaboration.

In Sec.~2, we present the relevant theoretical expressions required
for our analysis. In Sec.~3, we give the QCD fit results in the PT
and APT approaches and compare the deviation of the APT results from
the PT ones with the experimental uncertainties. Summarizing remarks
are given in Sec.~4.

\section{Basic relations}
In this section, necessary expressions for our analysis are written
and outline how we use them. We follow the well known approach based
on the Jacobi polynomial expansion of structure functions. This
method of solution of the Dokshitzer-Gribov-Lipatov-Altarelli-Parisi
(DGLAP) evolution equation \cite{DGLAP-1a,DGLAP-1b,DGLAP-2,DGLAP-3}
was proposed in Refs.~\cite{PS-1,PS-2} and developed for both
unpolarized
\cite{24-1a,24-1b,24-2a,24-2b,24-3,24-4,Sid-HT-96,Kat-HT-96} and
polarized cases \cite{LSS-1a,LSS-1b,LSS-2}. The main formula of this
method allows an approximate reconstruction of the structure
function through a finite number of Mellin moments
\begin{equation}
xF_{3}^{N_{max}}(x,Q^2)=\frac{h(x)}{Q^2}+x^{\alpha}(1-x)^{\beta}
\sum_{n=0}^{N_{max}} \Theta_n ^{\alpha , \beta}
(x)\sum_{j=0}^{n}c_{j}^{(n)}{(\alpha ,\beta )} M_{3}(j+2, Q^{2}) \,.
\label{Jacobi}
\end{equation}
Here $h(x)/Q^2$ is the HT term, $\Theta_n^{\alpha,\beta}$ are the
Jacobi polynomials, $c_j^{(n)}(\alpha,\beta)$  contain $\alpha$- and
$\beta$-dependent Euler $\Gamma$-functions where $\alpha,\beta$ are
the Jacobi polynomial parameters, fixed by the minimization of the
error in the reconstruction of the structure function.

The perturbative renormalization group $Q^2$ evolution of moments is
well known (see, e.g.,
Refs.~\cite{Buras:1979yt,Yndurain,Krivokhizhin:2009zz}) and in the
leading order reads as
\begin{equation} \label{evol-PT}
M^{pQCD}_3(N,Q^2) =\frac{~[\alpha _{s}(Q^{2})]^{\nu}} {~[\alpha
_{s}(Q_{0}^{2})]^{\nu}}\, M_3(N,Q^2_0) ,~~
{\nu(N)}={{{\gamma_{NS}^{(0),N}}/{2\beta_0}}}, ~N = 2,~3, ... \,,
\end{equation}
where $\alpha_s(Q^2)$~ is the QCD running coupling,
~$\gamma_{NS}^{(0),N}$ are the nonsinglet  one-loop anomalous
dimensions, $\beta_0=11-2n_f/3$ is the first coefficient of the
renormalization group $\beta$-function, and $n_f$ denotes the number
of active flavors.

Unknown quantity $M_3(N,Q^2_0)$ in Eq.~(\ref{evol-PT}) could be
parameterized as the Mellin moments of structure function $xF_3$ at
some point, $Q^2_0$:
\begin{equation}
M_3(N,Q^2_0)=\int_{0}^{1}dx{x^{N-2}}xF_3(x,Q_0^2)=
\int_{0}^{1}dx{x^{N-2}}Ax^{a}(1-x)^{b}(1+\gamma x), ~ N = 2,~3, ...
\,. \label{Mellin}
\end{equation}

The shape of the function $h(x)$ in Eq.~(\ref{Jacobi}) as well as
the parameters $A$, $a$, $b$, $\gamma$ in Eq.~(\ref{Mellin}), and
the scale parameter $\Lambda_{QCD}$ are found by fit of a combined
set of $xF_3$-data
\cite{CDHS,SCAT,BEBCGARGAMELLE,BEBC-WA59,NuTeV,CHORUS}. The detailed
description of the fitting procedure could be found in
Ref.~\cite{KKPS2}. The target mass corrections are taken into
account to the order $o(M^4_{nucl}/Q^4)$. We do not take into
account the nuclear effect in the $xF_3$ (see estimations of this
effect in Refs.~\cite{Tokarev-1,Tokarev-2}).

In our analysis we use the analytic approach in QCD, the APT
\cite{apt96-7-1,apt96-7-2,ShSol06}. This approach gives the
possibility of combining the renormalization group resummation with
correct analytic properties in $Q^2$-variable for physical
quantities. It should be noted that the moments of the structure
functions should be analytic functions in the complex $Q^2$ plane
with a cut along the negative real axis (see Ref.~\cite{ShSol06} for
more details), the ordinary PT description violates analytic
properties.
In the framework of the APT, the analytic coupling has no unphysical
singularities and in the low-$Q^2$ domain, instead of rapidly
changing $Q^2$ evolution, as occurs in the PT case, the APT approach
leads to rather slowly changing functions (see, e.g.,
Refs.~\cite{Khandramai:2011zd,GLS-our}). In the asymptotic region of
large $Q^2$ the APT and PT results coincide.

\begin{figure}
\begin{center}
{\includegraphics[width=8.0cm]{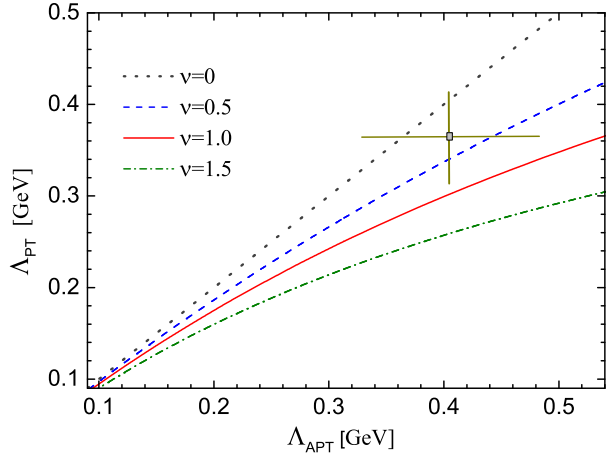}}\\
\caption {  \small { Values of $\Lambda_{{\rm PT}}$ vs.
$\Lambda_{{\rm APT}}$ from Eq.~(\ref{L_1}) for different values of
$\, \nu \,$. The point corresponds to extracted values of
$\Lambda_{{\rm PT}}$ and $\Lambda_{{\rm APT}}$ (with errors)
 presented in Table~1.}}
\label{Graph1}
     \end{center}
\end{figure}

In the framework of the APT, expression (\ref{evol-PT}) transforms
as follows:
\begin{equation}
{\cal{M}}^{APT}_3(N,Q^2)
 = \frac{{\cal{A}}_{\nu}(Q^{2})}
{{\cal{A}}_{\nu}(Q_{0}^{2})} \, {\cal{M}}_3(N,Q^2_0) \, ,
\label{evol-APT}
\end{equation}
where the analytic function ${\cal{A}}_{\nu}$ is derived from the
spectral representation and corresponds to the discontinuity of the
$\nu$th power of the PT running coupling
\begin{equation}
\label{del_APT} {\cal{A}}_{\nu}(Q^2) \,=\,\frac{1}{\pi}\,
\int_0^\infty d\sigma \,\frac{{\rm Im} \; \left \{ {\alpha}_{\rm
PT}^{\nu}(-\sigma -{\rm i}\varepsilon)
 \right\}}{\sigma\,+\,Q^2} \,.
\end{equation}
The mathematical tool for numerical calculations of
${\cal{A}}_{\nu}$ up to the four-loop level is given in
Ref.~\cite{KB13}. Note that the function ${\cal{A}}_{{\nu}=1}(Q^2)$
defines the APT running coupling, ${ \alpha}_{\rm APT}(Q^2)$.
In the leading order ($LO)$, the `normalized' analytic function
$\bar{\cal{A}}_{\nu}= [\beta_0  /(4 \pi)]^\nu {\cal{A}}_{\nu}$ has
rather a simple form (see Refs.~\cite{BMS05,Stefanis:2009kv}) and
can be writhen as
\begin{eqnarray}
 && \bar{\cal{A}}_{\nu}^{LO}(Q^2)=\left[\bar{a}^{LO}_{\rm PT}(Q^2)\right]^{\nu}\,- \,
\frac{{\rm {Li}_\delta}(t)}{\Gamma(\nu)} \, , \label{a_nu}  \\[0.2cm]
 && ~ {\rm {Li}}_{\delta}(t)= \sum_{k=1}^{\infty}
 \frac{t^k}{k^{\delta}} \, , ~~ t=\frac{\Lambda^2}{Q^2} \, ,~~~\delta=1-\nu \, ,
 \nonumber
\end{eqnarray}
where the `normalized' PT running coupling $\bar{a}^{LO}_{\rm
PT}(Q^2)=\beta_0 {{\alpha}_{\rm
PT}^{LO}(Q^2)}/(4\pi)=1/\ln(Q^2/\Lambda^2)\,$ and Li$_\delta$ is the
polylogarithm function.  For the case ${\nu=1}$ expression
(\ref{a_nu}) leads to the well-known APT result~\cite{apt96-7-1}
\begin{equation}
\label{a_1}
 \alpha_{\rm APT}^{LO}(Q^2)={\alpha}_{\rm {PT}}^{LO}(Q^2) +
 \frac{4 \pi}{\beta_0}\, \frac{\Lambda^2}{\Lambda^2-Q^2} \, .
\end{equation}
One could see that at large $Q^2$ the second term in the R.H.S. of
expression (\ref{a_1}) is negative. It was confirmed qualitatively
in the phenomenological analysis of the $xF_3$ data in
Ref.~\cite{sid1Q2}.
It should be stressed that values of the QCD scale parameter
$\Lambda$ are different in the PT and APT approaches. In order to
illustrate this feature, in Fig.~\ref{Graph1}, we present the
connection between the $\Lambda_{{\rm PT}}$ and the $\Lambda_{{\rm
APT}}$ following from the condition
\begin{equation}
\label{L_1} \left[ \alpha_{\rm PT}^{LO}(Q^2,\Lambda_{{\rm PT}})
\right]^{\nu} = {\cal{A}}_{\nu}^{LO}(Q^2, \Lambda_{{\rm APT}})
\end{equation}
at different powers of $\, \nu $. In this figure, the point marks
the $\Lambda_{{\rm PT}}$ and $\Lambda_{{\rm APT}}$ values (see
Table~1 below) with the corresponding experimental errors, which was
found by fitting the $xF_3(x,Q^2)$ data.

\section{Numerical results}

\begin{figure}[htb]
         \begin{minipage}[b]{0.48\textwidth}
\centering\includegraphics[width=0.97\textwidth]{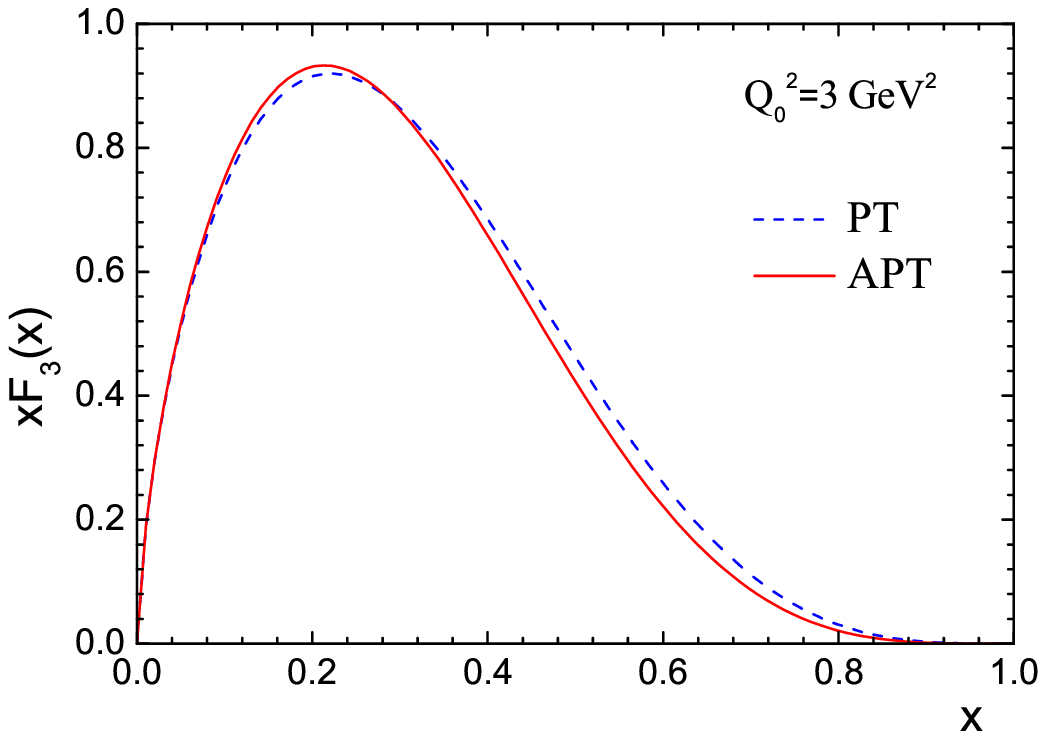}
         \end{minipage}
\phantom{}\hspace{0.3cm}%
     \begin{minipage}[b]{0.48\textwidth}
\centering\includegraphics[width=0.86\textwidth]{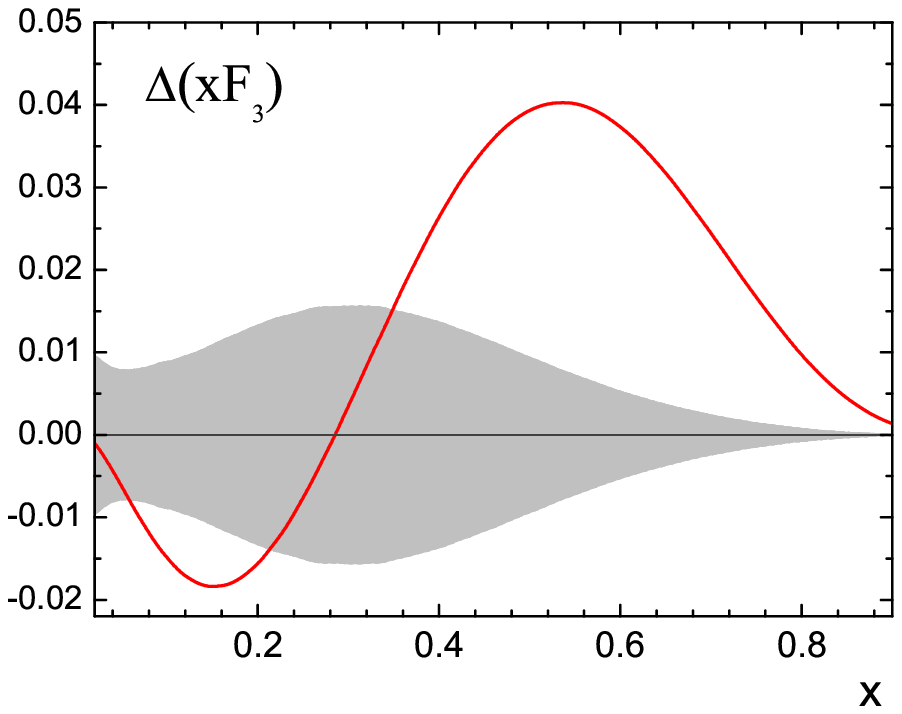} 
    \end{minipage}   \\[-1pt]
             \begin{minipage}[t]{0.45\textwidth}
\vspace*{-0.2cm}
\caption{ \small { The $x$-behaviour of $xF_3$
obtained in the APT (solid line) and the PT (dashed line) at
$Q_0^2=3$ GeV$^2$.}}
\vspace{0.2cm}
\label{Graph2}
        \end{minipage}%
\phantom{}\hspace{1.1cm}%
     \begin{minipage}[t]{0.45\textwidth}
\vspace*{-0.2cm} \caption{  \small { The difference $\Delta (xF_3)$
(solid line) from Eq.~(\ref{Del-xF3}) in comparison with the
experimental uncertainties of $xF_3^{APT}$ (shaded area) at
$Q_0^2=3$ GeV$^2$.}}
\vspace{0.2cm}
   \label{Graph3}
    \end{minipage}
    \end{figure}

\begin{figure}[htb]
         \begin{minipage}[b]{0.49\textwidth}
\centering\includegraphics[width=0.9\textwidth]{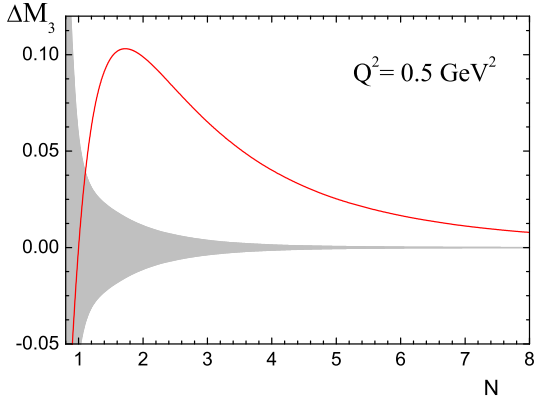}
         \end{minipage}%
\phantom{}\hspace{0.2cm}%
     \begin{minipage}[b]{0.49\textwidth}
\centering\includegraphics[width=0.9\textwidth]{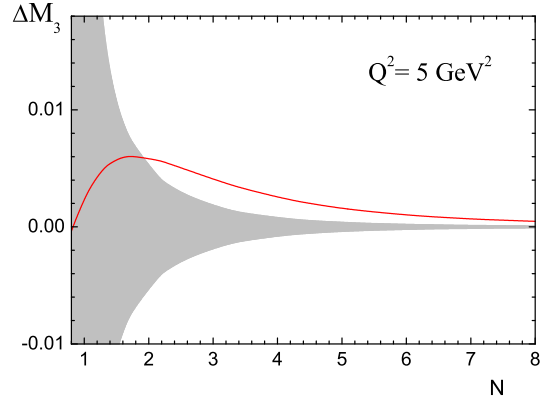} 
    \end{minipage}\\
    \vspace*{-0.2cm}
\caption{  \small {The difference $\Delta M_3(N,Q^2)$  (solid line)
from Eq.~(\ref{Delm3q2}) for $Q^2=0.5~{\mbox{\rm GeV}}^2$ (left
panel) and $Q^2=5~{\mbox{\rm GeV}}^2$ (right panel) is shown as a
function of a number of Mellin moment $N$. The shaded area shows the
experimental uncertainties of
 ${\cal{M}}^{APT}_3(N,Q^2)$.}}
\label{Graph4}
   \end{figure}

\begin{figure}
         \begin{minipage}[b]{0.49\textwidth}
\centering\includegraphics[width=0.9\textwidth]{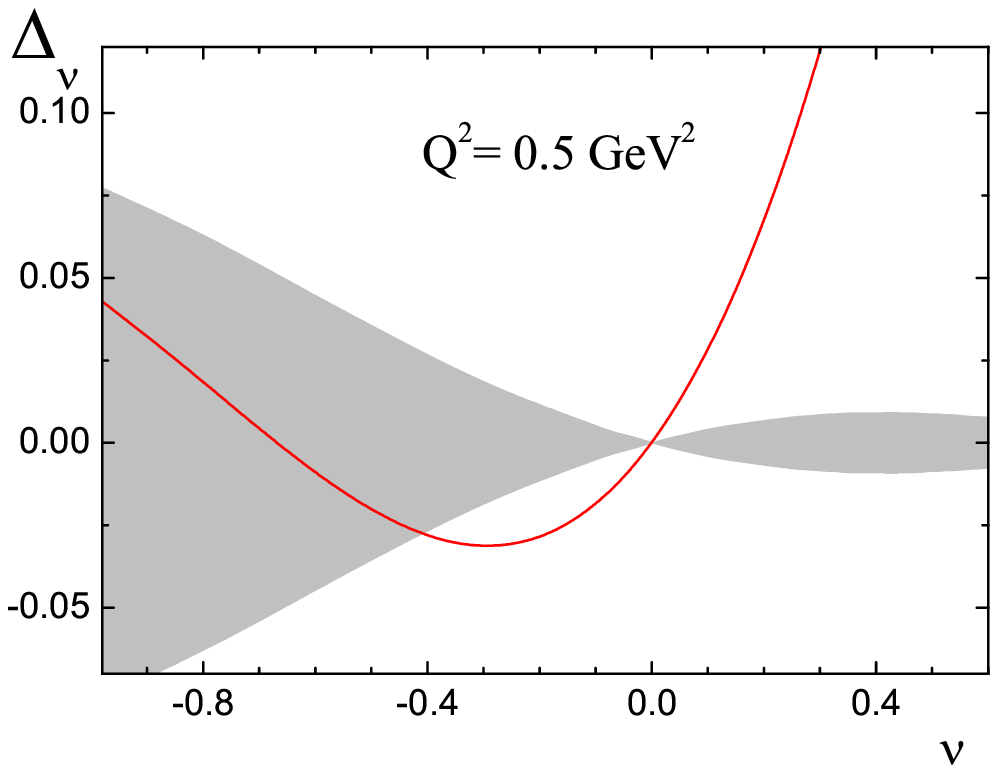}
         \end{minipage}%
\phantom{}\hspace{0.2cm}%
     \begin{minipage}[b]{0.49\textwidth}
\centering\includegraphics[width=0.9\textwidth]{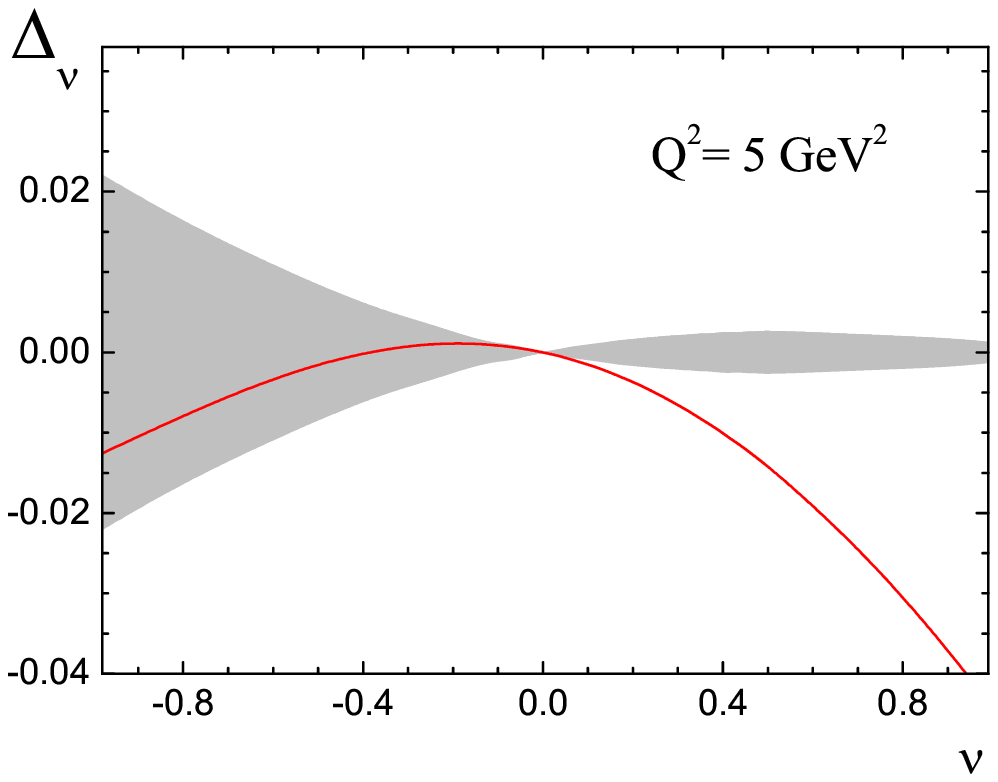} 
    \end{minipage}\\
    \vspace*{-0.4cm}
\caption{ \small {The quantity  $\Delta_\nu$ from Eq.~(\ref{Del-nu})
for $Q^2=0.5~{\mbox{\rm GeV}}^2$ (left panel) and $Q^2=5~{\mbox{\rm
GeV}}^2$ (right panel) compared with the experimental uncertainties
of the APT evolution factor ${\rm{E}}_{\nu}^{APT}$ (shaded area) at
$Q_0^2=3$~GeV$^2$.}} \label{Graph5}
   \end{figure}

\begin{figure}[h]
\phantom{}\vspace{1.0cm}%
     \begin{minipage}[b]{0.75\textwidth}
     \phantom{}\hspace{2.0cm}%
\centering\includegraphics[width=0.65\textwidth]{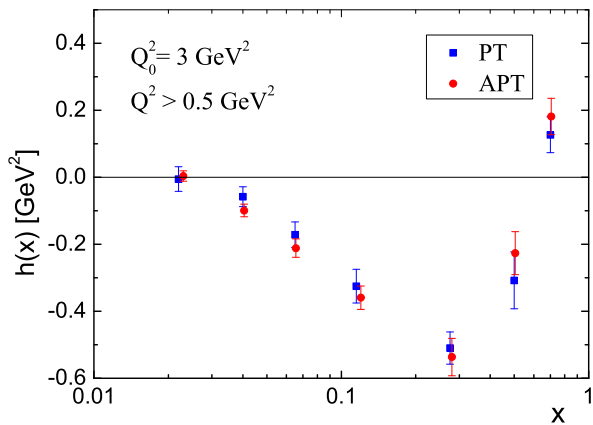} 
    \end{minipage}   \\
\vspace*{-0.2cm} \caption {\small {Higher twist contribution
resulting from the leading order QCD analysis of the $xF_3$ combined
data for the PT (squares) and APT (circles) approaches.} }
\label{Graph6}
   \end{figure}

The results of the leading order QCD fit of a combined set of
$xF_3$-data in the PT and APT approaches are presented in Table~1
and Figs.~\ref{Graph1} -- \ref{Graph6}. In both cases the PT and APT
are considered within the kinematic conditions $Q^2_0=3$~GeV$^2$,
$Q^2 > 0.5$~GeV$^2$, for number of active flavours $n_f=4$ and
$N_{Max}=11$. The values of $A$, $a$, $b$, $\gamma$, $h(x)$ and
$\Lambda$ are considered as free parameters.

\begin{table}[!t] \small
 \caption{\small {The results for the QCD leading order
fit of a combined set on the $xF_3$ data in the standard PT and the
APT approaches at $Q^2_0=3$~GeV$^2$, $Q^2 > 0.5$~GeV$^2$, $n_f=4$,
and $N_{Max}=11$. }}
\begin{center}\label{tab:1}
 {\begin{tabular}{|l|c|c|}   \hline
  & PT            & APT \\ [1mm]  \hline  \hline
~~ ~~ A        &  ~~~$3.75\pm1.97 $~~~   &  ~~~$3.74\pm0.64$~~~  \\
~~ ~~ $\alpha$   &  ~~~$0.65 \pm0.14 $~~~   &  ~~~$0.65\pm0.05$~~~  \\
~~ ~~ $\beta$    &  ~~~$3.62\pm0.07 $~~~   &  ~~~$3.97\pm0.04$~~~  \\
~~ ~~ $\gamma$   &  ~~~$2.75\pm2.33 $~~~   &  ~~~$3.56\pm0.89$~~~
\\ \hline ~~ ~$\Lambda$~[MeV]  &   ~~~$363 \pm49$~~   &  ~~~$407
\pm75$~~~   \\ \hline ~~~ ~ $\chi^2_{d.f.}$  &  ~~~$1.80$~~~   &
~~~$1.83$~~~   \\ \hline ~~~~ ~$x$  & $h(x)$~[GeV$^2$] &
$h(x)$~[GeV$^2$] \\ [1mm]  \hline  \hline
~~  $x<0.02$ & $-0.053\pm0.037$ &  $~~0.037\pm0.015$ \\
~~ $0.02-0.05$ & $-0.058\pm0.030$ & $-0.099\pm0.019$  \\
~~  $0.05-0.08$ & $-0.171\pm0.038$ &  $-0.211\pm0.027$ \\
~~  $0.08-0.15$ & $-0.325\pm0.050$ &  $-0.360\pm0.035$ \\
~~  $0.15-0.40$ & $-0.510\pm0.048$ &  $-0.537\pm0.056$ \\
~~ $0.40-0.60$ & $-0.308\pm0.085$ & $-0.227\pm0.064$  \\
~~  ~$x>0.60$ & ~~$0.127\pm0.053$ & ~~$0.181\pm0.054$  \\
\hline
\end{tabular} } \end{center}
\end{table}

As seen from Table~1, the analysis of a combined set of the
$xF_3$-data allows us to determine with good accuracy values of the
parameter $\Lambda_{QCD}$. These values are different for the PT and
APT cases, although consistent within errors. Errors in Table~1
correspond to $\Delta{\chi}^2=1$. It should be noted that due to the
increased volume of experimental data, in the present analysis the
errors for $\Lambda_{\rm PT}$ and $\Lambda_{\rm APT}$ are some times
less than in our previous analysis of CCFR data\cite{SS-F3-1} in
which it was obtained $\Lambda_{\rm PT}=363\pm 170$ MeV and
$\Lambda_{\rm APT}=350\pm 145$~MeV. Also, one can see from Table~1
that the highest twist contribution is defined much more precisely
than in our previous analysis \cite {SS-F3-1} (for comparison see
Fig.~4 in Ref.~\cite {SS-F3-1} and Fig.~\ref{Graph6} below).

Let us turn now to Fig.~\ref{Graph1}, in which the extracted
$\Lambda$-values is shown as a point with the corresponding
experimental errors. One could see this point is in the region of
small powers of $\, \nu\,$, $\nu< 1$, which corresponds effectively
to small numbers of the Mellin moments and, consequently, to small
values of $x$.

Figure 2 shows the $xF_3$-shape obtained in the APT (solid line) and
the PT (dotted line) cases. One can see that the result for the APT
approach is slightly higher than for the PT one for small $x$, and
less for large $x$.

Further, we will consider the difference between the PT and APT fit
results in comparison with the experimental errors and will try to
answer the question: In what kinematic areas the deviation goes
beyond the experimental uncertainties of the APT-fit result?

In Fig.~\ref{Graph3}, we plot the difference in the $x$-shape of the
$xF_3$ structure function
\begin{equation} \label{Del-xF3}
\Delta (xF_3)=xF_3^{PT}(x) - xF_3^{APT}(x) \,
\end{equation}
in comparison with the experimental uncertainties for $xF_3^{APT}$
at $Q_0^2=3$ GeV$^2$. Gray color shows a corridor of experimental
errors for $xF_3^{APT}$.
As can be seen from Fig.~\ref{Graph3}, the difference $\Delta
(xF_3)$ is beyond the experimental errors as at small $0.05<x<0.2$
as well at large values of a variable $x$, $x \gtrsim 0.4$. At large
$x$ values the deviation of the results of the PT from the APT
exceeds several times the width of a corridor of experimental
errors.

Figure \ref{Graph4} shows the difference between the moments in the
PT and APT determined by using Eqs.~(\ref{evol-PT}) and
(\ref{evol-APT}), respectively,
\begin{equation} \label{Delm3q2}
\Delta M_3(N,Q^2)=M^{PT}_3(N,Q^2) - {\cal{M}}^{APT}_3(N,Q^2) \,
\end{equation}
as a function of the number of the Mellin moments (including non
integral value of $N$) in a wide range of $\, 0.5< N<8$ at values
$Q^2=0.5~{\mbox{\rm GeV}}^2$ (left panel) and $Q^2=5~{\mbox{\rm
GeV}}^2$ (right panel). Comparing the results shown on the left and
right panels in Fig.~\ref{Graph4} one can see that in case of small
$Q^2$-values the difference $\Delta M_3$ goes beyond experimental
errors at $N\simeq 1$ and exceeds several times the width of this
corridor of the experimental errors at $N \gtrsim 1.5$. For
$Q^2=5~{\mbox{\rm GeV}}^2$ this effect is less pronounced.

Let us consider the difference between the PT and APT  approaches
for the evolution factor, ${\rm{E}}_{\nu}(Q^{2},Q_0^{2})$, which is
given by ${\rm{E}}_{\nu}^{PT}=[\alpha
_{s}(Q^{2})/\alpha_{s}(Q_{0}^{2})]^{\nu}$ for the PT  and
${\rm{E}}_{\nu}^{APT}={{\cal{A}}_{\nu}(Q^{2})}/
{{\cal{A}}_{\nu}(Q_{0}^{2})}$ for the APT, see Eqs.~(\ref{evol-PT})
and (\ref{evol-APT}), respectively.
In this case we consider the quantity
\vspace{0.1cm}
\begin{equation} \label{Del-nu}
\Delta_{\nu}(Q^2) = {\rm{E}}_{\nu}^{PT}(Q^2)  -
{\rm{E}}_{\nu}^{APT}(Q^2) \,
\vspace{0.1cm}
\end{equation}
as a function of noninteger $\nu$ at $Q^2_0=3$~GeV$^2$.
We present $\Delta_\nu(Q^2)$ in Fig.~\ref{Graph5} for values
$Q^2=0.5~{\mbox{\rm GeV}}^2$ (left panel) and $Q^2=5~{\mbox{\rm
GeV}}^2$ (right panel). The shaded area shows the experimental
uncertainties of ${\rm{E}}_{\nu}^{APT}(Q^2)$. This uncertainty is
related with errors of the only parameter $\Lambda_{{\rm APT}}$ (see
Table~1).


In general, the consideration of Figs.~\ref{Graph2} -- \ref{Graph5}
shows that the greatest difference between the results of the PT and
APT approaches should be expected at low $Q^2$ and large numbers of
Mellin moments $N$, which corresponds to large values of the
variable $x$.

Finally, in Fig.~\ref{Graph6} the results of the $x$ dependence for
the higher twists contribution in the PT and APT approaches are
presented. They are in qualitative agreement with each other. Our
results are in qualitative agreement with the results of other
papers on extraction of a contribution of the highest twists in the
structure function $xF_3$
\cite{Sid-HT-96,Kat-HT-96,Tokarev-1,Tokarev-2}.
Accuracy of definition of the function $h(x)$ considerably (a few
times) increased, in comparison with our result \cite{SS-F3-1} of
the analysis of CCFR data. In our previous analysis we noticed that
$h^{APT}(x) > h^{PT}(x)$ for $x>0.3$ \cite{SS-F3-1}. Here we confirm
this relation with higher accuracy. It should be noted that this
inequality is in qualitative agreement with the result obtained for
the shape of the HT contribution for the non-singlet approach for
the $F_2$ structure function \cite{KotKri10}. One could see from
Fig.~\ref{Graph6} that the opposite inequality takes place for small
values $x<0.3\,$: $h^{APT}(x) < h^{PT}(x)$.

\section{Conclusions}

This work could be considered as a continuation of our work
\cite{SS-F3-1}, in which for the first time the APT approach was
applied to the analysis of the $xF_3$ structure function data. In
this work, we used a combined set of data presented by various
collaborations. Thus, a new kinematic region of experimental data $
0.5~{\mbox{\rm GeV}}^2 <Q^2< 1.3$ GeV$^2$ was included in analysis.
As a result, the QCD scale parameter $\Lambda$, the parameters of
the shape of the structure function and the $x$-shape of the higher
twist contribution were defined with high accuracy. The shape of
this contribution is in qualitative agreement with the results of
the previous analysis of the $xF_3$ structure function data. We
compared the difference of the results of the PT and APT analysis
with the corridor of experimental uncertainties. We revealed that
the most essential difference in the PT and APT results should be
expected for small $Q^2$ and large numbers of Mellin moments, which
approximately corresponds to large values of a variable~$x$.

\section*{Acknowledgments}
It is a pleasure for the authors to thank S.~V.~Mikhailov,
O.~V.~Teryaev and V.~L. Khandramai for interest in this work and
helpful discussions.

This research was supported by the JINR-BelRFFR grant F14D-007, the
Heisenberg-Landau Program 2014, JINR-Bulgaria Collaborative Grant,
and by the RFBR Grants (Nos. 12-02-00613, 13-02-01005 and
14-01-00647).

\end{document}